%% file: main.tex
\documentclass[12pt]{article}
\usepackage{times}
\usepackage{geometry}
\usepackage[utf8]{inputenc}
\usepackage{enumitem,amssymb}

\usepackage{lineno}

\usepackage{hyperref}
\hypersetup{
    colorlinks=false,
    linkcolor=blue,
    filecolor=magenta,      
    urlcolor=blue,
}
 
\usepackage{siunitx}
\newcommand{\fermiLAT}{\emph{Fermi}-LAT}

\usepackage{graphicx}
\usepackage[version=3]{mhchem}

\usepackage{amsmath}

\usepackage{multirow}

\usepackage[table]{xcolor}

\usepackage{cleveref}
\renewcommand{\cref}{\Cref}

\title{All-sky Medium Energy Gamma-ray Observatory: Exploring the Extreme Multimessenger Universe}

\author{PI: Julie E. McEnery \\
NASA/GSFC\\~\\
(AMEGO Collaboration)\\~\\
\url{https://asd.gsfc.nasa.gov/amego/}
}

\begin{document}

\maketitle
\thispagestyle{empty}

\begin{abstract}

The All-sky Medium Energy Gamma-ray Observatory (AMEGO) is a probe class mission concept that will provide essential contributions to multimessenger astrophysics in the late 2020s and beyond. 
AMEGO combines high sensitivity in the 200 keV to 10 GeV energy range with a wide field of view, good spectral resolution, and polarization sensitivity. Therefore, AMEGO is key in the study of multimessenger astrophysical objects that have unique signatures in the gamma-ray regime, such as neutron star  mergers, supernovae, and flaring active galactic nuclei. 
The order-of-magnitude improvement compared to previous MeV missions also enables discoveries of a wide range of phenomena whose energy output peaks in the relatively unexplored medium-energy gamma-ray band.

\end{abstract}

\newpage
\thispagestyle{empty}
\input{coauthors}


\let\oldclearpage\clearpage
\cleardoublepage
\setcounter{page}{1}

\pagenumbering{arabic}

\include{execsummary}

\let\clearpage\relax
\include{science}

\include{technical}

\include{technology}

\include{schedule}

\include{cost}

\include{organization}



\let\clearpage\oldclearpage

\newpage

\bibliographystyle{unsrt}
\bibliography{main}

\end{document}

%% file: coauthors.tex

\begin{center}
\textbf{ The AMEGO Team Collaborators:}
\end{center}

\noindent Regina Caputo, 
S. Brad Cenko, 
Georgia De Nolfo, 
Alice Harding, 
Elizabeth Hays, \\ 
Julie McEnery (PI), 
John Mitchell, 
Jeremy Perkins, 
Judith Racusin, 
David Thompson, 
Tonia Venters, \textbf{(NASA/GSFC)};

\noindent Eric Burns, 
Carolyn Kierans, 
Lucas Uhm, 
Zorawar Wadiasingh, \textbf{(NPP/NASA/GSFC)};

\noindent Mattia Di Mauro, 
Alexander Moiseev, 
Elizabeth Ferrara, 
Sean Griffin, 
John Krizmanic, 
Amy Lien, 
Roopesh Ojha, 
Bindu Rani, 
Chris Shrader, 
Jacob Smith, \textbf{(CRESST/NASA/GSFC)};

\noindent Andrew Inglis, \textbf{(CUA/NASA/GSFC)};

\noindent Jessica Metcalfe, \textbf{(Argonne National Lab)};

\noindent Stefano Ciprini, 
Dario Gasparrini, 
Carlotta Pittori, \textbf{(ASI Space Science Data Center)};

\noindent Luca Zampieri, \textbf{(Astronomical Observatory of Padova)};

\noindent Aleksey Bolotnikov, \textbf{(Brookhaven National Lab)};

\noindent Brian Grefenstette, \textbf{(California Institute of Technology)};

\noindent Ulisses Barres, \textbf{(Centro Brasileiro de Pesquisas Fisicas)};

\noindent Jose-Manuel Alvarez, \textbf{(Centro De Laseres Pulsados)};

\noindent Marco Ajello, 
Dieter Hartmann, 
Lea Marcotulli, 
Lih-Sin The, \textbf{(Clemson University)};

\noindent Volker Beckmann, 
Denis Bernard, 
Jean-Philippe Lenain, \textbf{(CNRS/IN2P3)};

\noindent Christian Gouiffes, 
Isabelle Grenier, 
Philippe Laurent, \textbf{(Commissariat a l'Energie Atomique)};

\noindent Antonios Manousakis, \textbf{(Copernicus Astronomical Center)};

\noindent Vincent Tatischeff, \textbf{(CSNSM/IN2P3)};

\noindent Vaidehi S. Paliya, \textbf{(Deutsches Elektronen-Synchrotron (DESY))};

\noindent Joachim Kopp, 
Jan Lommler, 
Uwe Oberlack, \textbf{(Die Johannes Gutenberg-Universitaet Mainz)};

\noindent Naoko Kurahashi Neilson, \textbf{(Drexel University)};

\noindent Foteini Oikonomou, \textbf{(European Southern Observatory)};

\noindent Stefan Funk, \textbf{(Friedrich-Alexander-Universitaet Erlangen-Nuernberg)};

\noindent Cosimo Bambi, \textbf{(Fudan University)};

\noindent Sylvain Guiriec, 
Oleg Kargaltsev, 
Michael Moss, 
Alexander van Der Horst, 
George Younes, \textbf{(George Washington University)};

\noindent Nepomuk Otte, \textbf{(Georgia Tech)};

\noindent Daniel Castro, \textbf{(Harvard-Smithsonian CfA)};

\noindent Yasushi Fukazawa, 
Tsunefumi Mizuno, 
Masanori Ohno, 
Hiromitsu Takahashi, \textbf{(Hiroshima University)};

\noindent James Rodi, \textbf{(IAPS-INAF)};

\noindent Natalia Auricchio, \textbf{(INAF OAS Bologna)};

\noindent John B. Stephen, \textbf{(INAF/IASF Bologna)};

\noindent Elisabetta Bissaldi, 
Leonardo Di Venere, 
Francesco Giordano, 
M. Nicola Mazziotta, \textbf{(INFN Sezione di Bari)};

\noindent Sara Cutini, \textbf{(INFN Sezione di Perugia)};

\noindent Stefano Dietrich, \textbf{(Institute of Atmospheric Sciences and Climate)};

\noindent Manel Martinez , 
Javier Rico , \textbf{(Institut de Fisica d'Altes Energies (IFAE), The Barcelona Institute of Science and Technology (BIST))};

\noindent Ivan Agudo, 
Riccardo Campana, 
Martina Cardillo, 
Ezio Caroli, 
Stefano Del Sordo, 
Andrea Giuliani, 
Roberto Mignani, 
Antonio Stamerra, \textbf{(Instituto Nazionale di Astrofisica)};

\noindent Filippo D'Ammando, \textbf{(Istituto di Radioastronomia $\&$ INAF)};

\noindent Lukasz Stawarz, \textbf{(Jagiellonian University)};

\noindent Hidetoshi Kubo, \textbf{(Kyoto University)};

\noindent Jurgen Knodlseder, 
Luigi Tibaldo, \textbf{(L'Observatoire Midi-Pyrenees)};

\noindent Alexandre Marcowith, \textbf{(Laboratoire Univers et Particules de Montpellier)};

\noindent Peter Bloser, 
Chris Fryer, 
Pat Harding, 
Sam Jones, 
Alexei V. Klimenko, 
Hui Li, 
Lucas Parker, 
Richard Schirato, 
Karl Smith, 
Tom Vestrand, \textbf{(Los Alamos National Lab)};

\noindent Gottfried Kanbach, 
Andy Strong, \textbf{(Max Planck Institute for Extraterrestrial Physics)};

\noindent Kazuhiro Nakazawa, 
Hiro Tajima, \textbf{(Nagoya University)};

\noindent Michelle Hui, 
Daniel Kocveski, 
Colleen Wilson-Hodge, \textbf{(NASA/MSFC)};

\noindent Teddy Cheung, 
Justin Finke, 
J. Eric Grove, 
Matthew Kerr, 
Michael Lovellette, 
Richard Woolf, 
Eric Wulf, \textbf{(Naval Research Lab)};

\noindent Joseph Gelfand, \textbf{(New York University)};

\noindent Markus Boettcher, \textbf{(North West University South Africa)};

\noindent Maria Petropoulou, \textbf{(Princeton)};

\noindent Haocheng Zhang, \textbf{(Purdue University)};

\noindent Matthew Baring, \textbf{(Rice University)};

\noindent Matthew Wood, 
Eric Charles, 
Seth Digel, \textbf{(SLAC National Accelerator Laboratory)};

\noindent Vladimir Bozhilov, \textbf{(Sofia University)};

\noindent Manuel Meyer, 
Igor Moskalenko, 
Nicola Omodei, 
Elena Orlando, 
Troy Porter, 
Giacomo Vianello, \textbf{(Stanford University)};

\noindent Tim Linden, \textbf{(Stockholm University)};

\noindent John Beacom, \textbf{(The Ohio State University)};

\noindent S. Kaufmann, \textbf{(Universidad Autonoma de Chiapas)};

\noindent Miguel A. Sanchez-Conde, 
Juan Abel  Barrio, 
Alberto Dominguez, 
Marcos Lopez, 
Daniel Morcuende, \textbf{(Universidad Complutense de Madrid)};

\noindent Sonia Anton, \textbf{(Universidade de Aveiro)};

\noindent Rui Curado da Silva, \textbf{(Universidade de Coimbra)};

\noindent Stephan Zimmer, \textbf{(Universitaet Innsbruck)};

\noindent Martin Pohl, \textbf{(Universitaet Potsdam)};

\noindent Sara Buson, \textbf{(Universitaet Wurzburg Lehrstuhl fur Astronomie)};

\noindent Margarita Hernanz, 
Marc Riba, \textbf{(Universitat de Barcelona)};

\noindent Enrico Bozzo, 
Roland Walter, \textbf{(Universite de Geneve)};

\noindent Michael De Becker, \textbf{(Universite de Liege)};

\noindent Inga Stumke, \textbf{(Universitetet i Bergen)};

\noindent Silvia Zane, \textbf{(University College London)};

\noindent Michael Briggs, \textbf{(University of Alabama Huntsville)};

\noindent John Tomsick, 
Andreas Zoglauer, \textbf{(University of California Berkeley)};

\noindent Steven Boggs, \textbf{(University of California San Diego)};

\noindent Robert Johnson, 
David Williams, \textbf{(University of California Santa Cruz)};

\noindent Jamie Holder, \textbf{(University of Delaware)};

\noindent Pablo Saz Parkinson, \textbf{(University of Hong Kong)};

\noindent Brian Fields, 
Xilu Wang, \textbf{(University of Illinois)};

\noindent Markos Georganopoulos, 
Eileen Meyer, \textbf{(University of Maryland Baltimore County)};

\noindent Peter Shawhan, \textbf{(University of Maryland College Park)};

\noindent Bing Zhang, \textbf{(University of Nevada Las Vegas)};

\noindent Fabian Kislat, 
Marc McConnell, 
Chanda Prescod-Weinstein, \textbf{(University of New Hampshire)};

\noindent Jack Hewitt, \textbf{(University of North Florida)};

\noindent Eugenio Bottacini, 
Michele Doro, 
Luca Foffano, \textbf{(University of Padova)};

\noindent Denis Bastieri, 
Alessandro De Angelis, 
Elisa Prandini, 
Riccardo Rando, 
Luca Baldini, 
Barbara Patricelli, \textbf{(University of Pisa $\&$ INFN)};

\noindent Francesco Longo, \textbf{(University of Trieste $\&$ INFN)};

\noindent Stefano Ansoldi, \textbf{(University of Udine)};

\noindent Wlodek Bednarek, \textbf{(Uniwersytet Lodzki)};

\noindent Jim Buckley, 
Wenlei Chen, 
Henric Krawczynsiki, \textbf{(Washington University in St. Louis)};

\noindent Harsha Blumer, \textbf{(West Virginia University)};

\noindent Paolo Coppi, \textbf{(Yale University)};

%% file: execsummary.tex
\section{Executive Summary}
 The All-sky Medium Energy Gamma-ray Observatory (AMEGO) is a probe class mission that will provide ground-breaking new capabilities for multimessenger astrophysics - identifying and studying the astrophysical objects that produce gravitational waves and neutrinos. AMEGO also has compelling science drivers in astrophysical jets, compact objects, dark matter and nuclear line spectroscopy. AMEGO will cover the energy range from 200 keV to over 10 GeV, with more than an order of magnitude improvement in sensitivity relative to previous missions. 

\begin{figure}[!h]
\begin{minipage}[c]{0.60\textwidth}
\includegraphics[width=\textwidth]{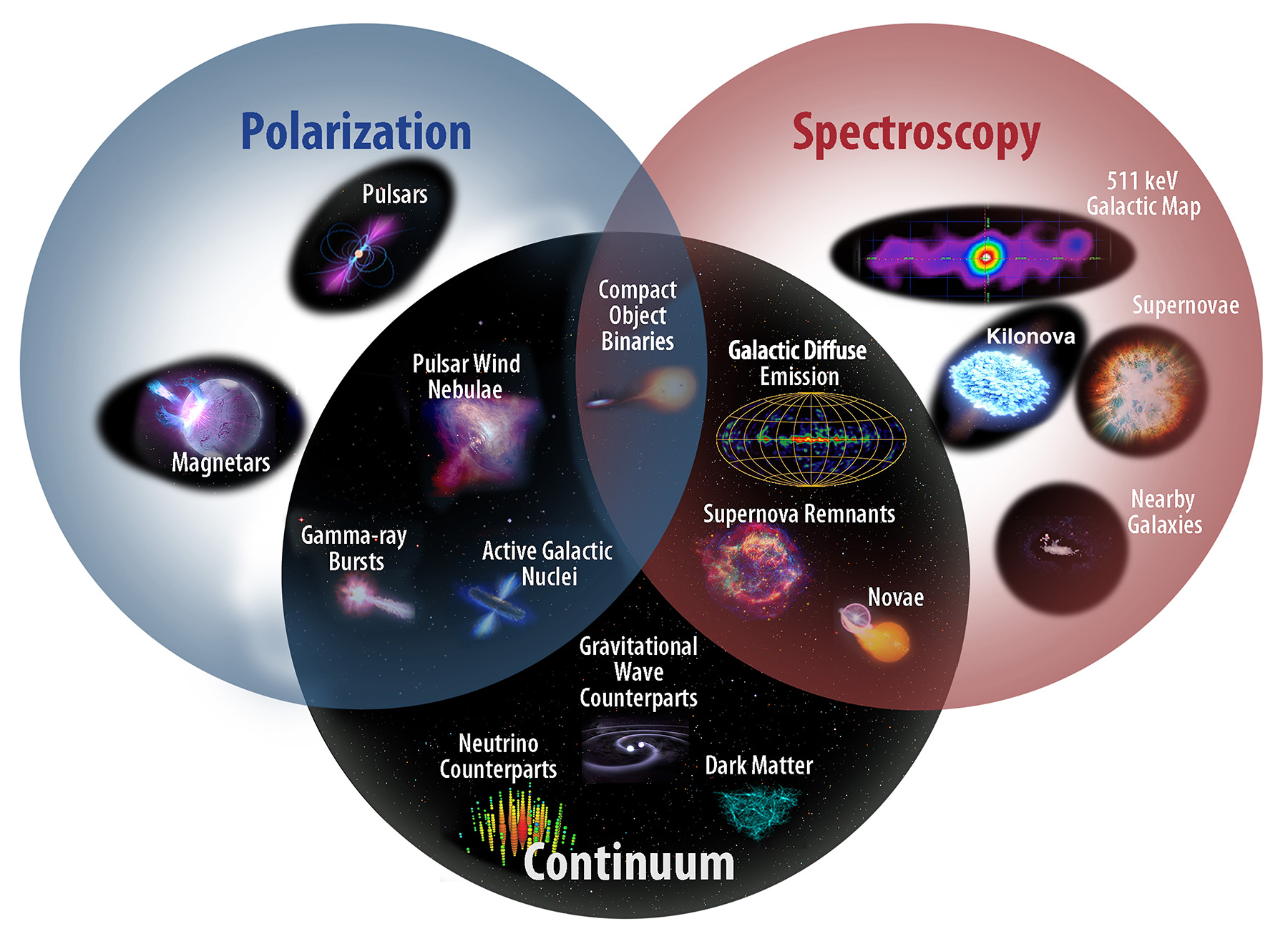}
\end{minipage}\hfill
\begin{minipage}[c]{0.40\textwidth}
\caption{\small \it AMEGO will provide breakthrough capabilities in three areas of MeV astrophysics: a wide field of view and broad energy range will provide outstanding capability in time-domain and multimessenger astrophysics with excellent synergies with observations at other wavelengths; polarization capability will uniquely probe conditions and processes in astrophysical jets and in the magnetospheres and winds of compact objects; nuclear line spectroscopy will bring new insight into the topical area of element formation in dynamic environments.
\label{fig:AMEGOVenn}}
\end{minipage}
\vspace{-4mm}
\end{figure}

\noindent{\bf Ground-Breaking Capabilities:} Developments in detector technology since the last major mission in medium energy gamma-ray astrophysics enable a transformative probe class mission. 

\vspace{-0.2cm}
\begin{table}[!ht]
\centering
\small
\rowcolors{1}{blue!9!white!97!green}{blue!6!white!98!green}
\caption{\small \it  AMEGO's design has been optimized for excellent flux sensitivity, broad energy range, and large field of view. AMEGO will have $\sim$5x better angular resolution below 200 MeV than Fermi-LAT. Its good energy resolution, combined with large effective area, will enable ground-breaking nuclear line spectroscopy. Additionally, AMEGO will make polarization measurements below 5 MeV.}
\vspace{2mm}
\begin{tabular}{|c|p{10.5cm}|}
\hline
{\bf Energy Range} & 200 keV -- $>$10 GeV \\
{\bf Angular Resolution} & 2.5$^\circ$ (1 MeV), 1.5$^\circ$ (5 MeV), 2$^\circ$ (100 MeV)\\
{\bf Energy Resolution ($\sigma/$E)} & $<$$1\%$ ($<$2 MeV), $\sim$ 10$\%$ (1 GeV) \\
{\bf Field of View} & 2.5 sr (20\% of the sky) \\
{\bf Line Sensitivity} & $1\times 10^{-6}$ ph cm$^{-2}$ s$^{-1}$ for the 1.8 MeV $^{26}$Al line in 5 years\\ 
{\bf Polarization Sensitivity} & $<$20\% MDP for a source 1\% the Crab flux, observed for 10$^6$ s \\
{\bf Sensitivity (MeV s$^{-1}$ cm$^{-2}$)} & $2\times10^{-6}$ (1 MeV), $1\times10^{-6}$ (100 MeV) in 5 years\\
\hline
\end{tabular}
\label{tab:properties}
\end{table}

\noindent {\bf Community and partnerships:} The AMEGO team is an international group of 200 scientists at 80 institutions with extensive experience designing, building, and operating gamma-ray telescopes. 

\vspace{0.2cm}
\noindent {\bf Mature Technology:} The technologies used in AMEGO are mature, and we have developed and tested key hardware and analysis technologies with support from agencies in the US and Europe. The AMEGO subsystems and spacecraft have undergone preliminary engineering and costing studies that show that this mission is build-able within the probe class cost envelope.

%% file: science.tex
\section{Key Science Goals \& Objectives}\label{sec:science}


 The under-explored gap of medium-energy gamma rays has rich scientific promise. In the last decade, high-energy satellites have provided exquisite observations of the sky in both the hard X-ray/low-energy gamma-ray (E$\sim$10-200 keV) and high-energy gamma-ray (E$>$ 1 GeV) bands. We now know that some of the most extreme astrophysical objects have peak emission in the MeV band~\cite{Paliya:2019oyn,2019arXiv190305648W,Ojha:2019xan}. 
 Only MeV gamma-ray observations allow for nuclear spectroscopy, providing the only direct view of nuclear processes in supernovae and kilonovae~\cite{2019arXiv190202915F}. Finally, the MeV band covers the positron annihilation line at 511 keV,
 where the source of Galactic positrons is one of the prevailing mysteries in MeV astrophysics. 

Gamma-ray observations have played a critical role in every multimessenger source identified to date -- including gamma-ray lines seen from SN1987A, a nearby neutrino source \cite{1988Natur.331..416M}; 
a gamma-ray burst from the neutron star merger event GW170817A \cite{2017ApJ...848L..12A}; 
and a gamma-ray flare from the active galaxy TXS 0506+056, the first identified counterpart to a high-energy neutrino source \cite{2018Sci...361.1378I}. 
In each of these cases, the gamma-ray observations were critical to understanding the underlying physical phenomena driving these extremely energetic sources~\cite{Burns:2019guq}.

AMEGO will provide capabilities more than an order of magnitude better than previous MeV gamma-ray missions, which, when combined with the upcoming generation of gravitational wave and neutrino observatories, will lead to a revolution in multimessenger astrophysics~\cite{Burns:2019zzo, 2019arXiv190505089K}.


\begin{figure}[!ht]
\begin{minipage}[c]{0.45\textwidth}
\includegraphics[width=\textwidth]{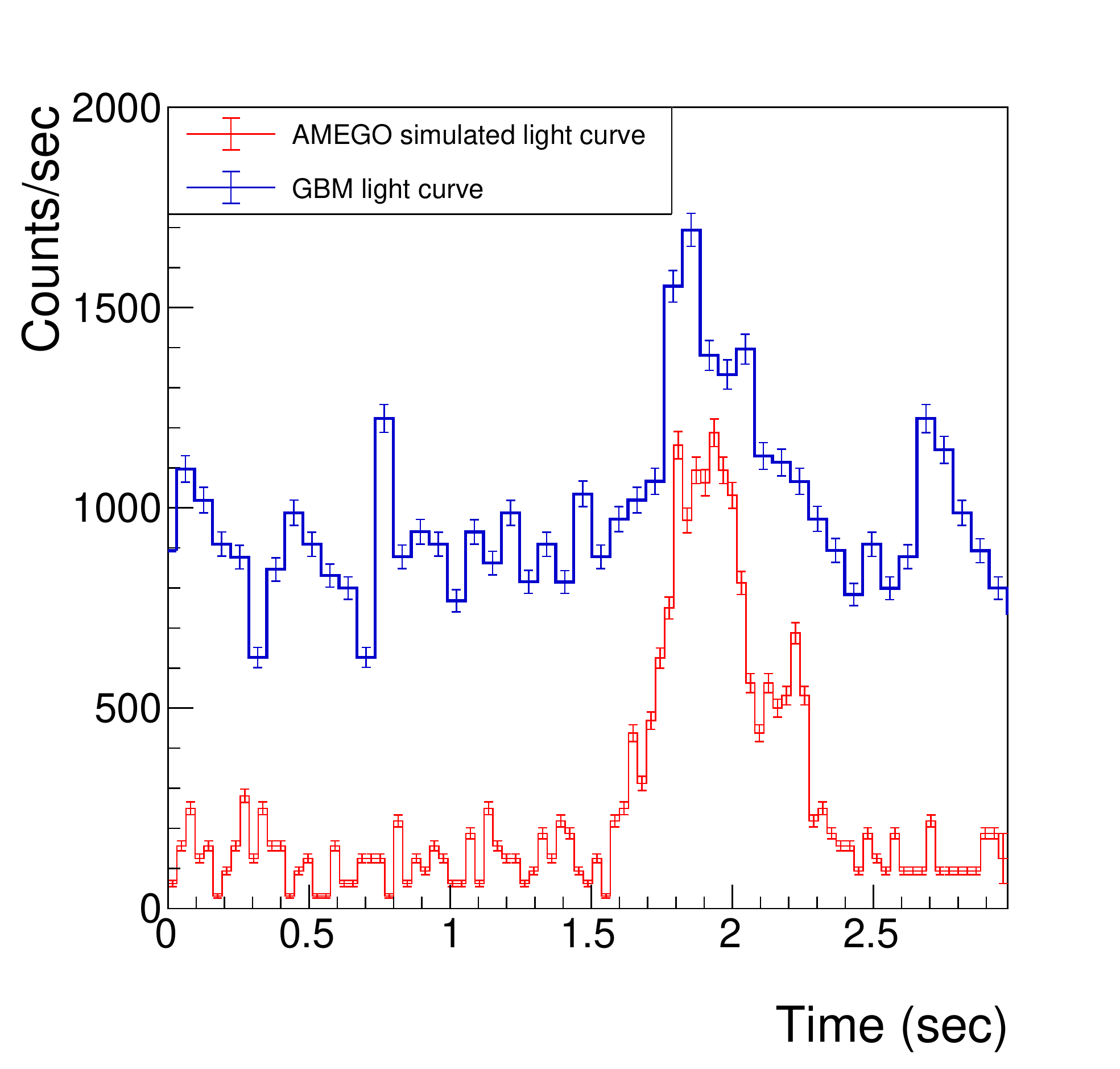}
\end{minipage}\hfill
\begin{minipage}[c]{0.55\textwidth}
\includegraphics[width=\textwidth]{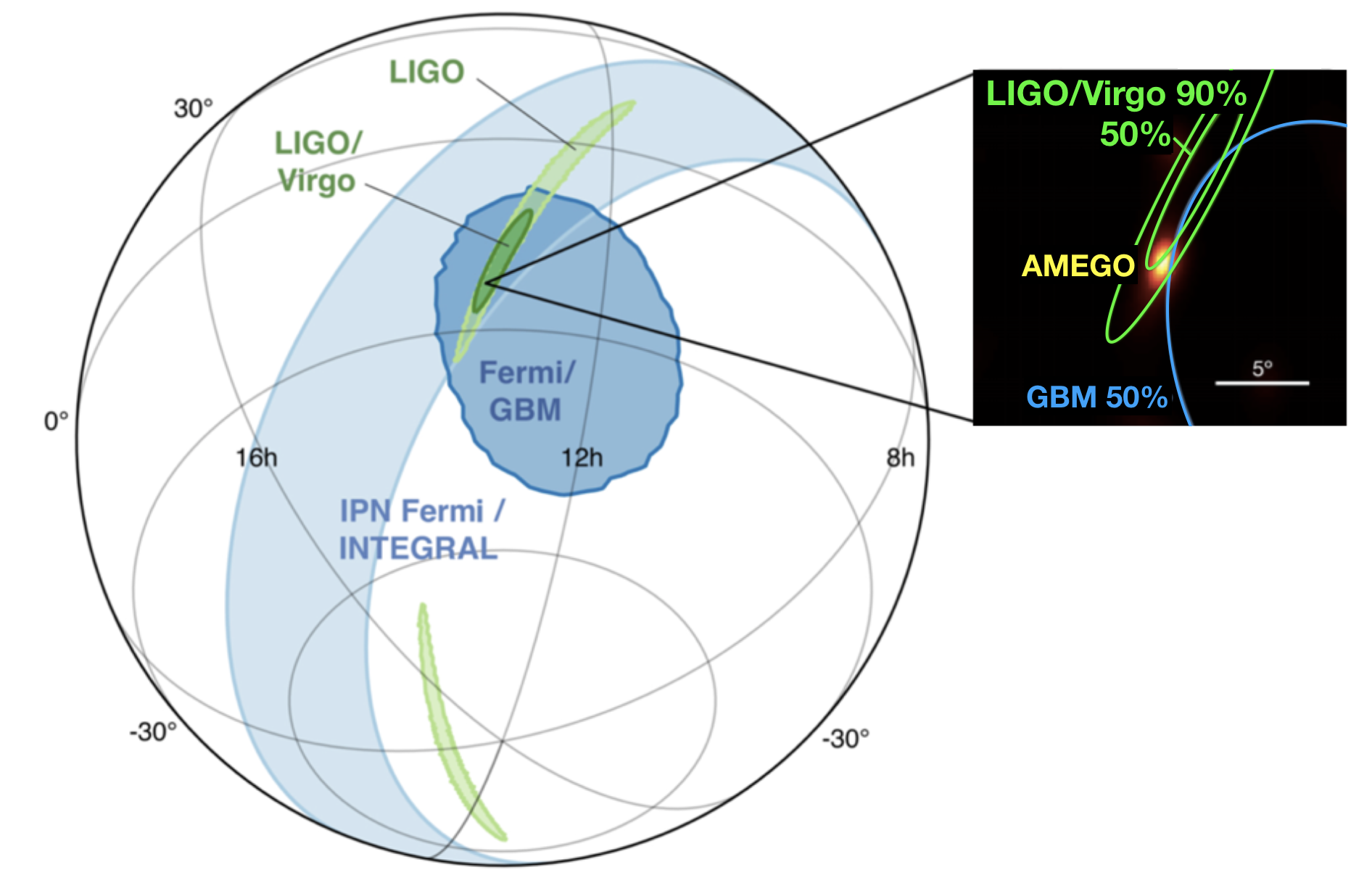}
\end{minipage}
\caption{\small \it AMEGO will detect GRB 170817A-like events with high signal-to-noise ratio due to excellent background rejection and large effective area. {\it Left:} The simulated AMEGO lightcurve (200--600 keV) and the observed GBM lightcurve (50-300 keV) for GRB 170817A. AMEGO could detect this burst out to 120 Mpc, giving a detection volume 25 times greater than Fermi-GBM. {\it Right:} The simulated AMEGO skymap for this event showing a statistical localization of 0.5$^{\circ}$ for the GRB.
}
\label{fig:170817Example}
\vspace{-4mm}
\end{figure}

The field of multimessenger astrophysics (the identification and study of astrophysical objects that produce gravitational waves, neutrinos and cosmic rays) has burst into prominence in the past few years. By the end of the next decade, it will no longer be sufficient to simply find and identify electromagnetic counterparts; the focus will be on leveraging joint electromagnetic and gravitational wave/neutrino/cosmic-ray observations to address compelling science questions on the nature of these extreme sources and using the unique multimessenger data as a probe of fundamental physics~\cite{Burns:2019zzo,2019arXiv190303582B}. The peak power output of electromagnetic counterparts lies in the gamma-ray band for the objects identified so far -- short gamma-ray bursts (GRB) as gravitational wave counterparts and blazars (active galaxies whose jets are aligned to our line of sight) for neutrino counterparts. 

\vspace{0.2cm}
\noindent{\bf Compact Objects, Gamma-ray bursts and Gravitational wave counterparts:} 
Neutron star (NS) mergers, which include both neutron star/neutron star and neutron star/black hole mergers, are the prototypical multimessenger events, producing gravitational waves followed by both short gamma-ray bursts and the nuclear-powered kilonova. Multimessenger observations of these cataclysmic events will probe sources of gravitational waves and astrophysical neutrinos, enable precision cosmology, and provide unique probes of fundamental physics, the origin of heavy elements, the behavior of relativistic jets, and the equation of state of supranuclear matter~\cite{2019arXiv190303582B}. 
AMEGO will provide crucial electromagnetic capabilities to understand these sources. AMEGO will detect an order of magnitude more well-localized short GRBs than any prior mission and will have a joint GW-GRB detection rate of order one per month. Through timing, spectral, and polarization observations of the prompt gamma-ray burst, AMEGO will test general relativity, probe the physical conditions within the ultrarelativistic jet, and provide a GW-GRB sample for tests of cosmology. For nearby events AMEGO can directly measure the nuclear production, providing a new test of kilonova models. The sub-degree localizations will enable prompt follow-up characterization of these sources, enabling a fuller understanding of merger remnants.


\vspace{0.2cm}
\noindent{\bf Element formation in extreme environments - kilonovae and supernovae:}
Medium-energy gamma rays provide a unique probe of astrophysical nuclear processes, directly measuring radioactive decay, nuclear de-excitation, and positron annihilation. The substantial information carried by gamma-ray photons allows us to see deeper into these objects; the bulk of the power is often emitted at gamma-ray energies; and radioactivity provides a natural physical clock that adds unique information~\cite{2019arXiv190202915F,2019arXiv190505089K,2019arXiv190503793W}.
The process of Galactic chemical evolution is driven by the star-formation rate and the relative rates of dynamic nucleosynthesis sources, predominantly thermonuclear (Type Ia) and core-collapse (Type II, Type 1b/c) supernovae (SNe), along with kilonovae resulting from compact object mergers.
AMEGO can probe Galactic evolution, interstellar medium dynamics, propagation of cosmic rays, and relativistic-particle acceleration through gamma-ray line spectroscopy measurements of short- and long-lived radioactive isotopes~\cite{2019arXiv190202915F}.

\begin{figure}[!ht]
\begin{minipage}[c]{0.47\textwidth}
\caption{\small  \it AMEGO observations will reveal the explosion geometry of SN Ia by measuring the temporal evolution of the nuclear line fluxes. This is illustrated with simulated AMEGO $^{56}$Co lightcurves for two different nickel distributions in the ejecta of SN2014J. The green data points are what were measured by SPI. The inset illustrates simulated background subtracted spectra for AMEGO observations 80 days after the explosion (for the $^{56}$Ni core model). AMEGO observations of SN Ia out to distances of $\leq$ 50 Mpc will provide a significant improvement in our understanding of the SN Ia progenitor system(s) and explosion mechanism(s).  \label{fig:SN1a}}
\end{minipage}
\begin{minipage}[c]{0.52\textwidth}
\includegraphics[width=\textwidth]{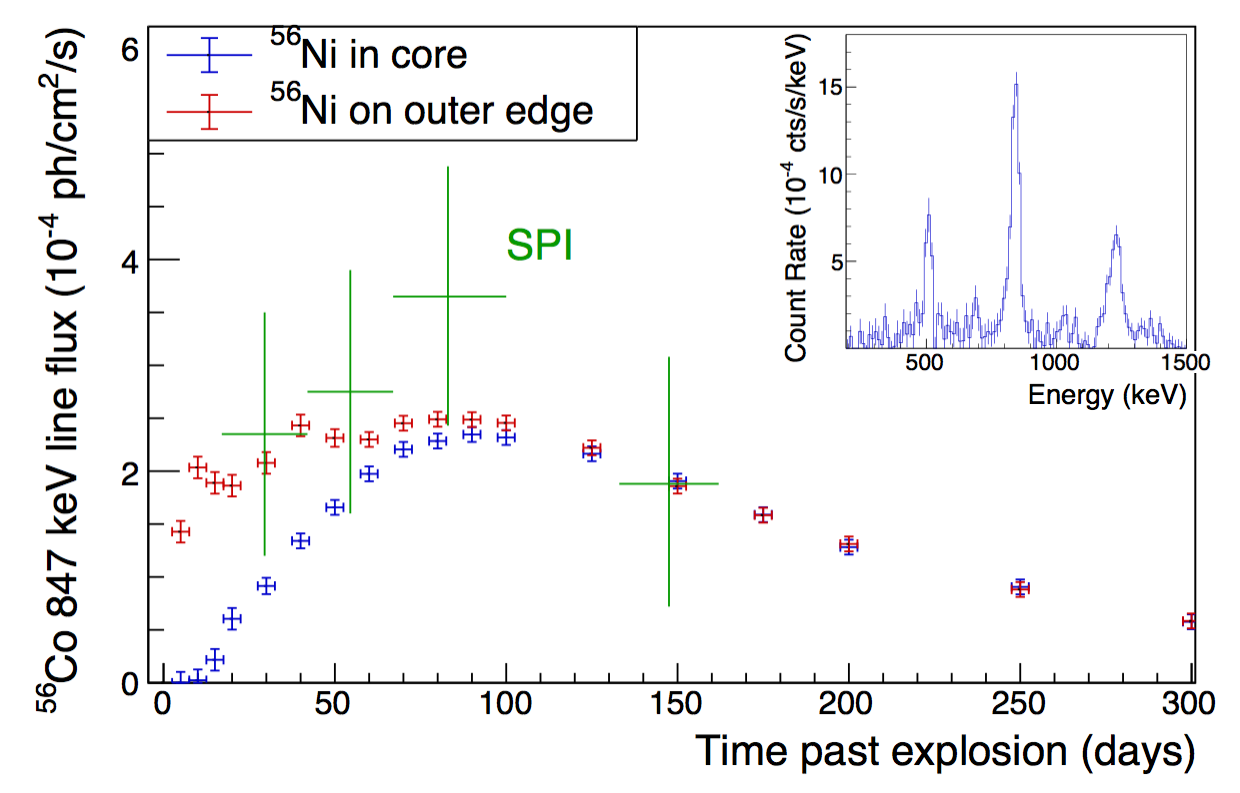}
\end{minipage}\hfill
\vspace{-4mm}
\end{figure}

The nature of the progenitors of SN Ia and how they explode remains elusive~\cite{2014ARA&A..52..107M}. The lack of a physical understanding of the explosion introduces uncertainty in the extrapolations of the properties of SN Ia to the distant universe.
Time-domain characterization of the emergent SN Ia gamma ray lines will help extract physical parameters such as explosion energy, total mass, spatial distribution of nickel masses, and ultimately lead to the astrophysical modeling and understanding of progenitors and explosion mechanisms~\cite{2019arXiv190202915F}.

\vspace{0.2cm}
\noindent{\bf Astrophysical jets - nature's extreme accelerators:}
All extragalactic sources of MeV gamma-rays are candidate neutrino and ultra-high energy cosmic-ray sources~\cite{Venters:2019cnk, 2018A&A...620A.174K, Orlando:2019sxa, 2018MNRAS.475.2724O}: AMEGO will detect over 500 long GRBs per year, and hundreds of blazars with peak power in the MeV band -- known to be the most luminous objects of their type. AMEGO observations (including polarization measurements) will be fundamental to constraining the composition of the high-energy particles in the jets of neutrino-emitting objects. These data, combined with neutrino observations by IceCube and KM3NET will identify and provide key insights to understanding nature's most extreme accelerators. AMEGO will provide essential observations and breakthrough science in partnership with the enhanced gravitational wave and neutrino observations expected by the end of the next decade.


\begin{figure}[!ht]
\begin{minipage}[c]{0.53\textwidth}
\includegraphics[width=\textwidth]{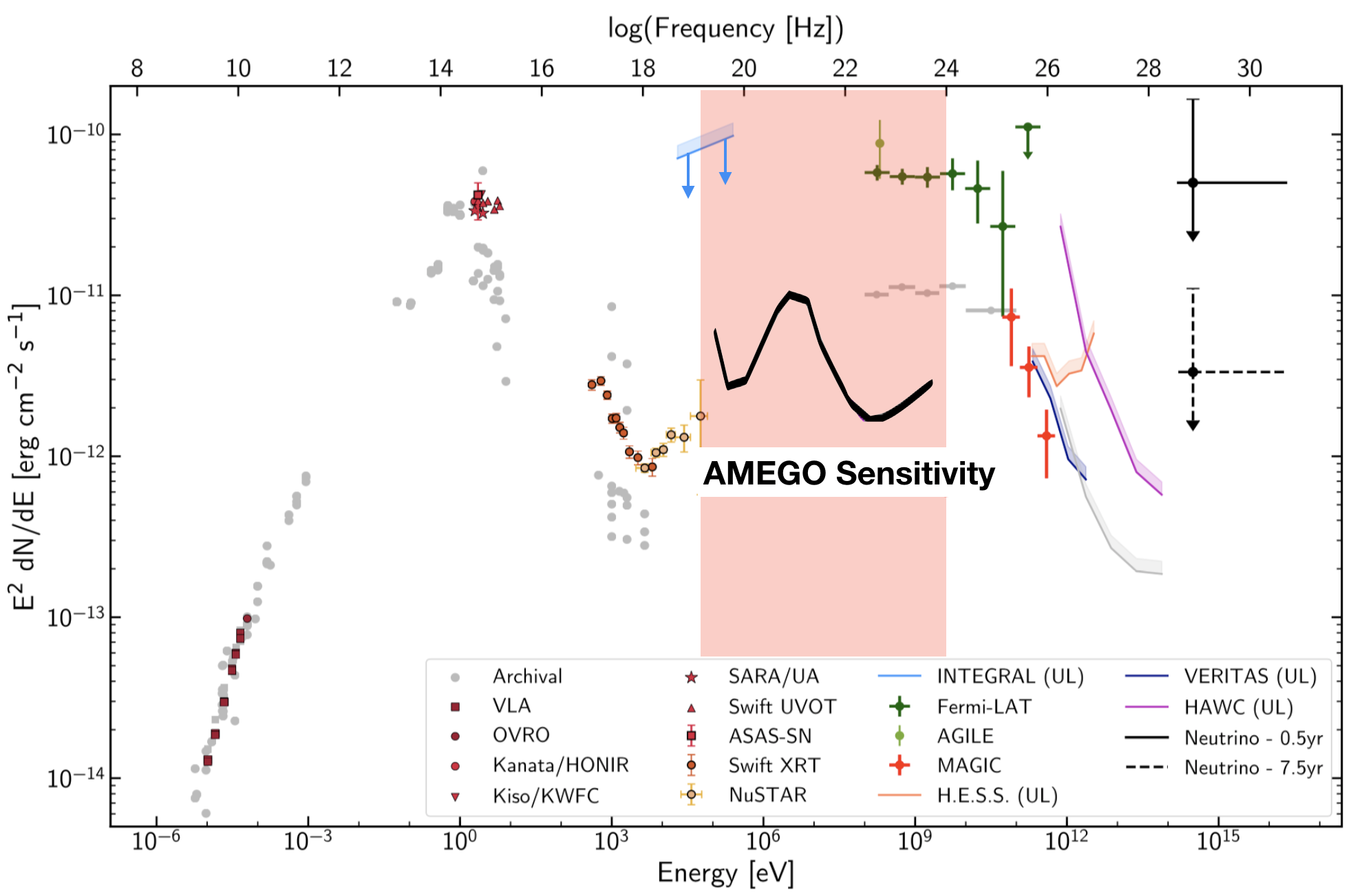}
\end{minipage}\hfill
\begin{minipage}[c]{0.45\textwidth}
\caption{\small \it The recent association ($\sim$ 3$\sigma$) of the blazar TXS 0506+056 with a high-energy neutrino lends support to the possibility that relativistic blazar jets in particular may be sources of gamma rays, neutrinos, and cosmic rays~\cite{2018Sci...361.1378I}. 
AMEGO will not only continue monitoring of these blazars currently provided by \fermiLAT~but will also provide the needed connection between the hard X-ray and the high-energy gamma-ray spectra which is known to be critical to understanding the particle content of these jets~\cite{Keivani:2018rnh}. 
\label{fig:TXS0506}}
\end{minipage}
\vspace{-4mm}
\end{figure}

The brightest and most powerful blazars (such as flat-spectrum radio quasars, FSRQs) tend to have their peak emission in the MeV band, making it the most effective band to probe radiation and particle acceleration in blazars. Improving information in the 1-100 MeV band will augment that from other wavelengths to enable separation of different radiation components~\cite{2019ApJ...874L..29R}. Polarization signals in the MeV band can distinguish between blazar emission models and also constrain the neutrino production from secondary pair synchrotron
and the gamma-ray radiation by primary proton synchrotron radiation~\cite{2019ApJ...876..109Z, 2019BAAS...51c.348R, Meyer:2019trp}. Furthermore, current theories suggest that the MeV band is the most important for constraining the neutrino production through the accompanying cascading pair synchrotron counterpart~\cite{Ojha:2019xan}.  
Medium-energy gamma-ray astronomy, enabled by technological advances that will be realized in the coming decade, will provide a unique and indispensable perspective on the jetted sources important to multimessenger astrophysics.

\vspace{0.2cm}
\noindent{\bf AMEGO as an observatory:}
In addition to the exciting multimessenger science described above, AMEGO will be a powerful general-purpose observatory providing substantial new discovery capability to the scientific community. 
Importantly, AMEGO will continue the long-term monitoring of the gamma-ray sky that the {\it Fermi} instruments have maintained for over a decade.  
In addition to watching the universe for transients like GRBs, monitoring can search for long-term trends, accumulate statistics for faint sources, and provide context for multimessenger discoveries as new facilities emerge \cite{2018Galax...6..117T, DiMauro:2019ujo}. 

\begin{figure} [!ht]
	\centering
     \includegraphics[width=.48\textwidth]{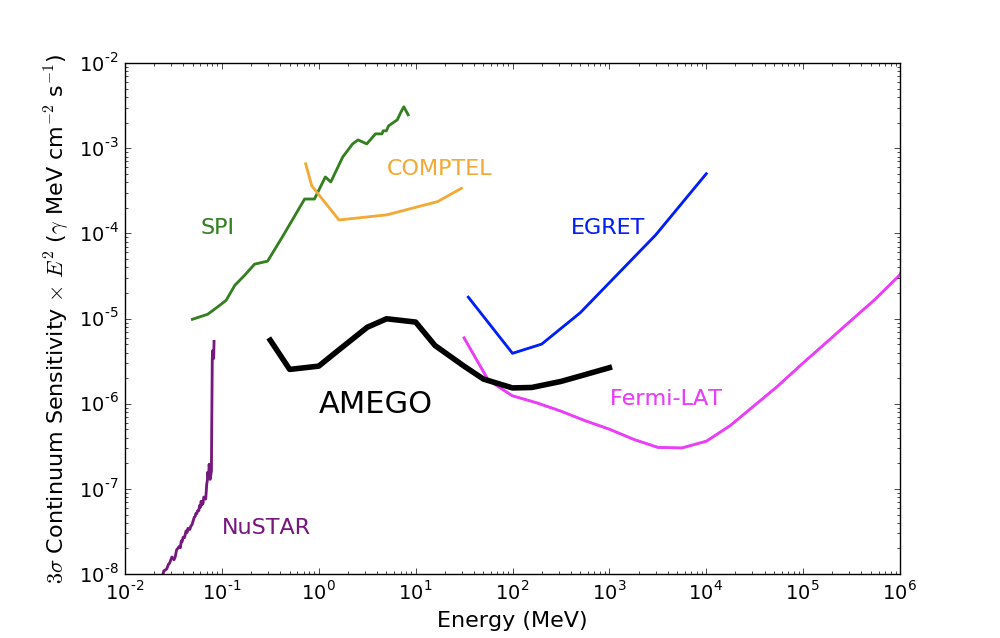}
     \includegraphics[width=.46\textwidth]{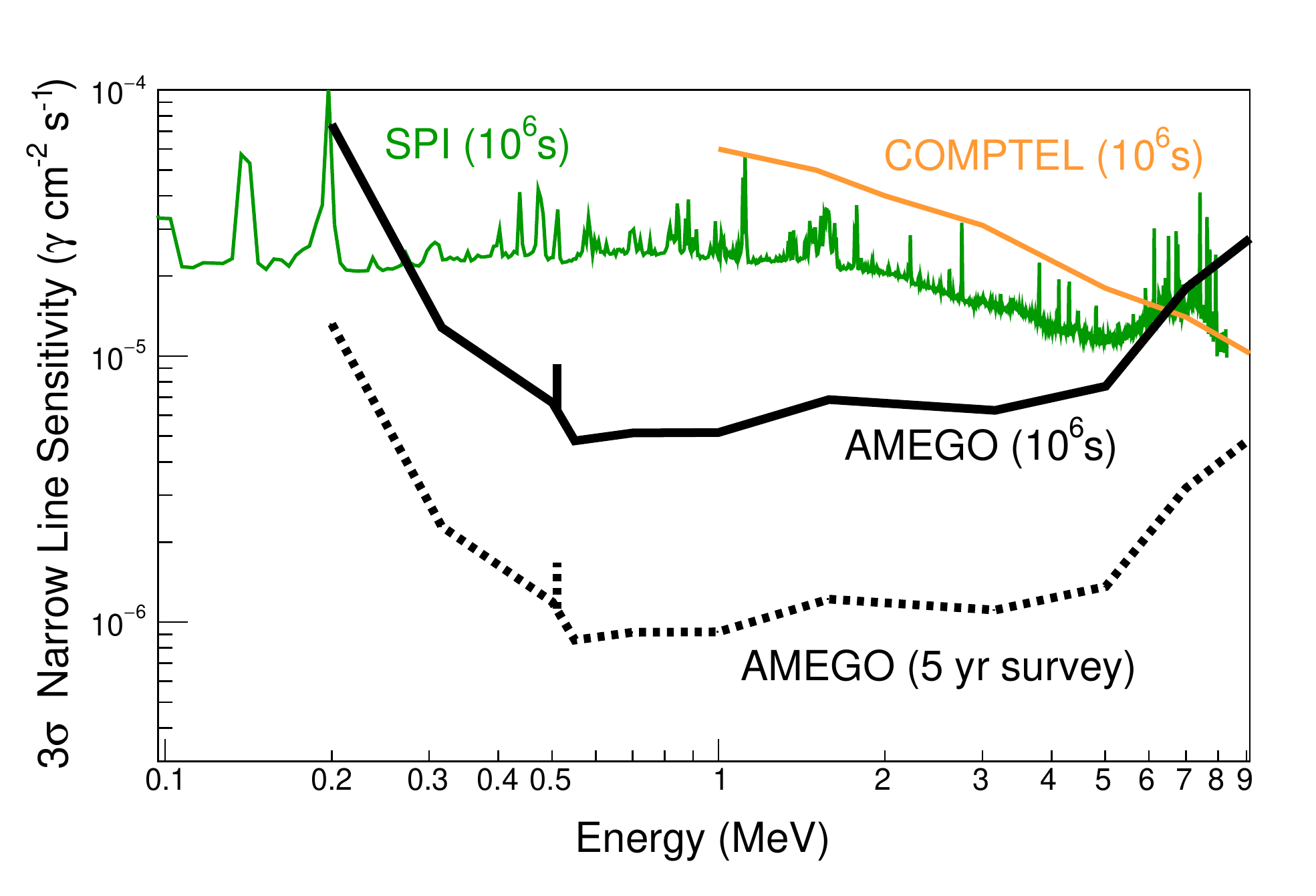}
     \caption{ \small \it 
     Left:
     AMEGO will revolutionize medium-energy gamma-ray astronomy by providing $>$20$\times$ better sensitivity in the MeV band compared to COMPTEL. The AMEGO 3$\sigma$ continuum sensitivity is calculated for all-sky exposure during a 5-year mission.  All other sensitivity curves are calculated for typical source exposure over the mission lifetimes.
     Right:
     Narrow-line sensitivity for AMEGO from 200~keV to 9~MeV, in comparison to INTEGRAL/SPI and COMPTEL. 
     }
     \label{fig:Sensitivity}
\end{figure}

AMEGO's discovery potential derives from its combination of order-of-magnitude improvement in continuum sensitivity, advanced nuclear line spectroscopy, and gamma-ray polarization capability~\cite{Rani:2019ber, 2019whitepaper}; its superior angular resolution will help resolve the diffuse gamma-ray emission that limits the sensitivity of Fermi-LAT analyses $<$300 MeV.  
Discovery is almost a foregone conclusion from this level of performance gain.  One target area for potential breakthroughs is magnetar physics.  
Magnetars are laboratories for extreme quantum electrodynamics, with critical spectral and polarization features in the AMEGO energy range that relate directly to issues in fundamental physics such as the QED prediction photon splitting and birefringence of the magnetized vacuum~\cite{2019arXiv190305648W}.  Another opportunity is the ongoing search for the nature of dark matter, with well-motivated possibilities potentially revealed in the medium-energy gamma-ray band \cite{2019arXiv190305845C, 2017JCAP...05..001B,2018PhRvD..98k6009K, 2016arXiv160803591F}.

%% file: technical.tex
\section{Technical Overview}\label{technical}

\begin{figure}[!ht]
\begin{minipage}[c]{0.5\textwidth}
\includegraphics[width=\textwidth, trim=15 100 15 60, clip=true]{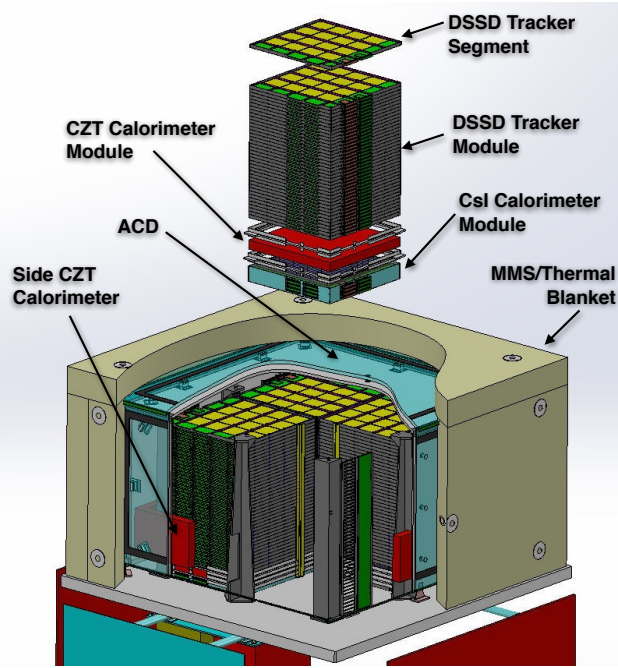}
\end{minipage}\hfill
\begin{minipage}[c]{0.45\textwidth}
\caption{\small \it A mechanical sketch of AMEGO highlights the four subsystems with a cutaway and one tower extracted. The 60 layer double sided silicon detector (DSSD) tracker converts or scatters incoming gamma rays and accurately measures the positions and energies of either the electron-positron pair or the Compton-scattered electron passing through the instrument. 
The CZT calorimeter modules sit beneath and cover the outer sides of the lower layers of the tracker modules. The CsI calorimeter modules consist of hodoscopic layers of crystal logs at the base of the instrument.  The modular design of 4 towers (each comprising tracker, CZT, CsI modules) sit within top and side panels of the ACD. A micrometeoroid shield (MMS) and thermal blanket cover the top and sides of the instrument. 
\label{fig:AMEGOExploded}}
\end{minipage}
\vspace{-4mm}
\end{figure}

To achieve the science described in {\bf Section~\ref{sec:science}} and in the numerous studies submitted to the Astro2020 white paper call, we optimized an instrument design that enables sensitive continuum spectral measurements over a broad energy range, with additional capabilities for measuring polarization and spectral lines. 
AMEGO consists of four subsystems: ({\bf Fig.~\ref{fig:AMEGOExploded}}) a double-sided silicon 
detector (DSSD) tracker, a 3D position-sensitive virtual Frisch-grid Imaging Cadmium Zinc Telluride (CZT) calorimeter, a segmented thallium-activated Cesium Iodide (CsI) calorimeter and a plastic scintillator Anti-Coincidence Detector (ACD). 
The Si tracker, CsI calorimeter, and ACD are analogs of systems flown on \fermiLAT~\cite{2009ApJ...697.1071A}. The AMEGO instrument performance derived from detailed detector simulations~\cite{2017ICRC...SIMS} is shown in {\bf Table \ref{tab:properties}}.


\vspace{0.2cm}
\noindent{\bf The tracker subsystem} consists of 60 layers of DSSDs. 
For the candidate design presented in this proposal, each tower layer contains a 4 $\times$ 4 array of DSSDs, each 9.5 cm square and 500 microns thick with 500 micron strip pitch. The DSSDs are wire bonded on the top and the bottom with x- and y-strips and read out on one x and y side~\cite{2019arXiv190209380G}. 
The layers are separated by 1.0~cm ({\bf Fig.~\ref{fig:AMEGOExploded}}). 
This configuration was used to calculate the sensitivity curve shown in {\bf Fig.~\ref{fig:Sensitivity}}. 

\vspace{0.2cm}
\noindent{\bf The CZT calorimeter} 
consists of an array of 8~mm $\times$ 8~mm $\times$ 40~mm bars. One layer sits below the tracker and another extends partially up the outer sides of the tracker ({\bf Fig.~\ref{fig:AMEGOExploded}})~\cite{2019HEAD...1710902H, 2017JInst..12C2037M}. 
It provides a precise measurement of the location and energy of the scattered gamma ray. 
The CZT bars are operated in a drift mode that enables 3-dimensional reconstruction of the location of the interaction in the detector, thus providing excellent positional resolution ($<$1 mm) as well as very good energy resolution ($<$ 1\% at 662 keV) at room temperature.  

\vspace{0.2cm}
\noindent{\bf The CsI(Tl) calorimeter} lies below the CZT and provides the depth to contain enough of a pair-conversion generated electromagnetic shower to extend the energy range to GeV energies. 
The design is similar to the \fermiLAT~CsI calorimeter, but it dramatically improves the low-energy performance by collecting the scintillation light with silicon photo-multipliers (SiPMs).
Each calorimeter module consists of 6 layers of CsI(Tl) crystal bars (15~mm$\times$15~mm$\times$380~mm) arranged hodoscopically ({\bf Fig.~\ref{fig:AMEGOExploded}})~\cite{2019arXiv190105828W}. 
The CsI(Tl) bars are wrapped in a reflective material to give high light collection efficiency, and the scintillation light is read out by an array of bonded SiPMs. 

\vspace{0.2cm}
\noindent{\bf The Anti-Coincidence Detector} is the first-level defense against the charged cosmic-ray background that outnumbers the gamma rays by 3--5 orders of magnitude. It covers the top and four sides of the tracker ({\bf Fig.~\ref{fig:AMEGOExploded}}). 
The ACD utilizes the same type of plastic scintillator as used on the LAT with wavelength shifting (WLS) strips and a SiPM readout. 
WLS strips are inserted in grooves in each panel edge and viewed by two SiPMs, allowing more uniform light collection than with SiPMs alone.

From detailed simulations using the MEGAlib toolkit~\cite{2006NewAR..50..629Z}, we can calculate the performance of the AMEGO instrument (\textbf{Table~\ref{tab:properties}}). 
The energy resolution is particularly important in the MeV regime where sources of gamma-ray line emission are prominent.
The angular resolution not only affects the quality of images, but aids in resolving source confusion and improves the sensitivity of the instrument. For details on the instrument simulations, please see Ref.~\cite{2017ICRC...SIMS}. 

\vspace{0.2cm}
\noindent\textbf{Operations concept:}
AMEGO will fly in a low-inclination, low-Earth orbit. AMEGO data will be downlinked through 10-12 TDRSS Ka band contacts per day. The science data will be downlinked to the operations center with a latency requirement of 12 hours (with a 3 hour latency goal). In addition, on board science processing will identify and localize gamma-ray bursts and send alerts to the science community in near real time.

Thanks to its large field-of-view ($\sim$2.5 sr) and sky-scanning mode, the instrument will observe the entire sky every 3 hours (two orbits). 
As a result, AMEGO will serve as a powerful all-sky detector for the transient and variable emission expected from extreme phenomena. 
In the era of time-domain (LSST, SKA, WFIRST) and multi-messenger (gravitational wave and neutrino observatories) astrophysics, AMEGO will be an essential contributor.

%% file: technology.tex
\section{Technology Drivers} \label{sec:technology}


The AMEGO concept builds on the strong heritage of the \fermiLAT~and technology developed for gamma-ray and cosmic-ray detectors. 
Most components are at or above TRL~6. 
The division of the smaller AMEGO instrument into four tower units instead of 16 reduces complexity in comparison to LAT by allowing the electronics to be placed on the outer sides of the detector.


The AMEGO design calls for double-sided silicon detectors (DSSDs) in the tracker instead of single-sided silicon strip detectors as used in LAT to achieve sufficient spatial resolution for the reconstruction of electron tracks from both sub-100 MeV pair-production and Compton scattering interactions. 
DSSD trackers have proven flight performance in PAMELA \cite{2003NIMPA.511...72A} and AMS-02 \cite{2008NIMPA.593..376A}. 
The DSSDs should be held in an open window-frame composite structure to eliminate passive material between layers and minimize the passive material contained within the tracker volume.
Laboratory, beamline, and suborbital testing of DSSDs and their electronic readout as part of the ComPair prototype will be done through a NASA-funded APRA (PI: J.~E. McEnery).

\begin{figure}[!ht]
\begin{minipage}[c]{0.40\textwidth}
\includegraphics[width=\textwidth]{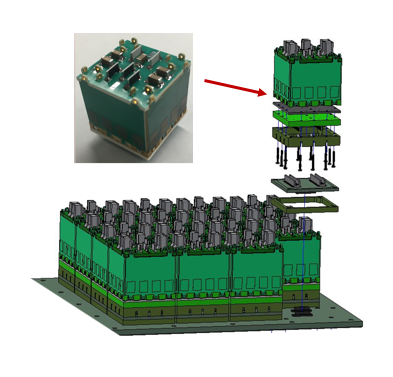}
\end{minipage}
\begin{minipage}[c]{0.55\textwidth}
\caption{\small \it 
Laboratory tests demonstrate that the CZT detectors can achieve good energy resolution at room temperature in the AMEGO energy range, placing this technology at TRL 4.
The module prototype, built with CZT bars, has been tested with various radioactive sources (e.g. \ce{^{232}U}, \ce{^{137}Cs}), as well as in the beam test at HIGS, and demonstrated expected performance.
The energy resolution measured after correcting for the calibrated 3-dimensional response was found to be (FWHM) $\sim$0.7$\%$ \cite{2015ITNS...62.3193B}. \label{fig:CZT}}
\end{minipage}
\vspace{-4mm}
\end{figure}

The low-power SiPM sensors 
selected for both the ACD and CsI calorimeter have successfully operated for the past 6 months on NRL’s SIRI experiment on STPSat-5, a spacecraft operated by the DoD Space Test Program~\cite{SIRI}.  These components and their accompanying readout are also the subject of two NASA-funded APRA projects, the Advanced Scintillator Compton Telescope (ASCOT) balloon project (PI: P.~F. Bloser) and a CsI calorimeter for the ComPair prototype (PI: J.~E. Grove).


Although CZT pixel detectors have been used in suborbital and orbital applications \cite{2013ITNS...60.4610H, 2005SSRv..120..143B}, the AMEGO CZT calorimeter is the lowest TRL subsystem as a deep CZT layer is required. 
This can be accomplished by using the CZT-bar technology being developed for ground-based applications at Brookhaven National Laboratory (BNL) \cite{2015ITNS...62.3193B}.
A NASA-funded APRA  project (PI: D.~J. Thompson) is underway to space-qualify CZT bars operated in a drift mode. 

The simulations and analysis algorithms required for AMEGO are either based on {\it Fermi} heritage or have been developed for the MEGA Compton and pair telescope \cite{2005NIMPA.541..310K} and for the NASA-funded COSI balloon payload (PI: S. Boggs) \cite{2015NIMPA.784..359C}. 
In addition, NASA has approved an APRA project to further improve the imaging technologies for future Compton telescopes (PI: A. Zoglauer).

%% file: schedule.tex
\section{Schedule}

\begin{figure}[!ht]
	\centering
	 \includegraphics[width=0.9\textwidth,trim=100 260 50 70, clip=true]{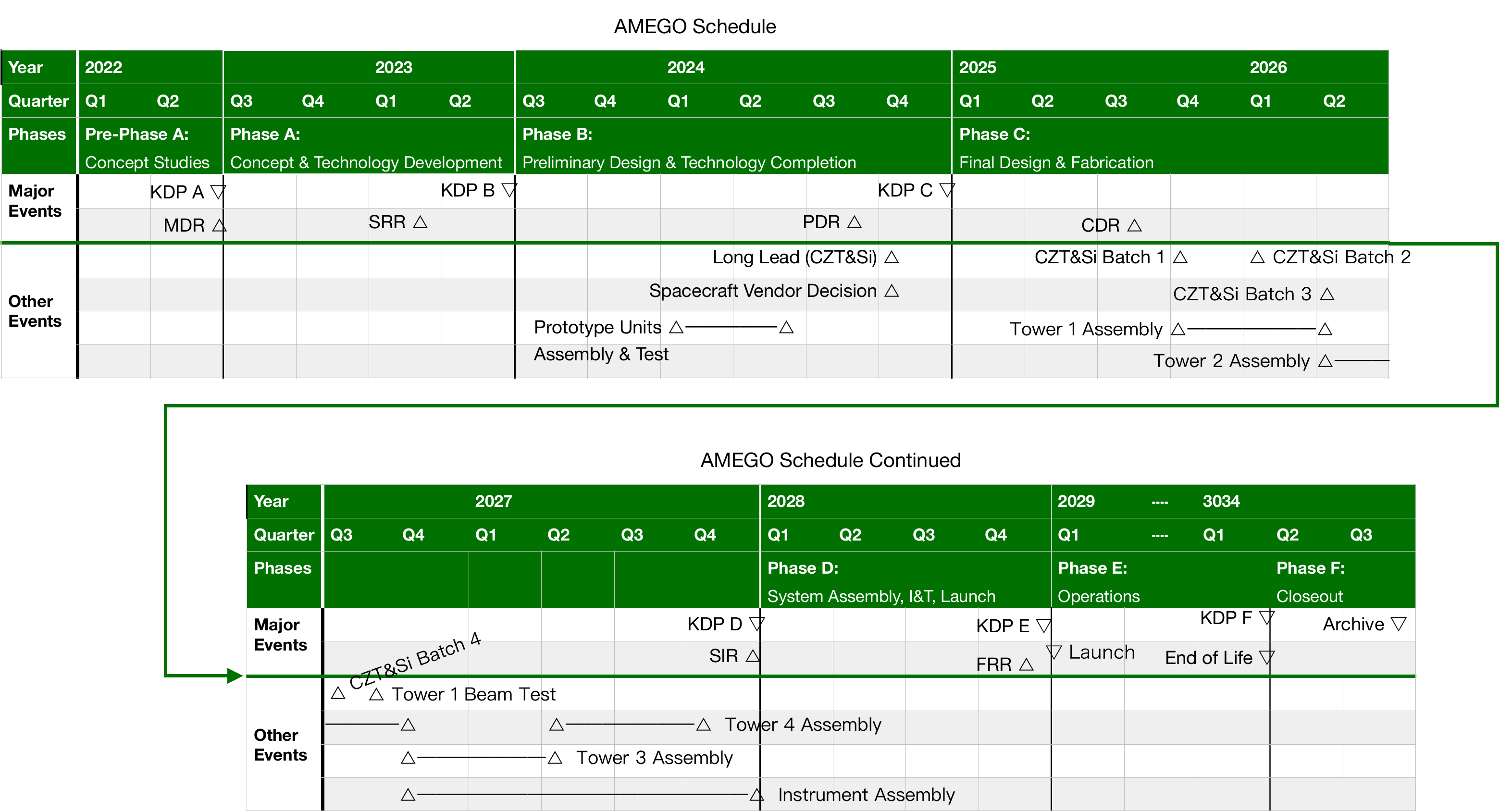}
     \caption{\small \it We have identified key milestones and assume approval to proceed occurs in early 2022.  
     Critical items are listed in other events including long-lead orders (Si and CZT) as well as selecting a spacecraft vendor.  
     Prototype units of the four subsystems are assembled and tested in Phase B.  The integration of the towers occur in stages and includes the Si Tracker, CsI and CZT Calorimeters (the ACD is not detailed).  Tower 1 will be calibrated in a beam while Tower 2 is being assembled.}  
     \label{fig:schedule}
\end{figure}

The AMEGO schedule (Figure \ref{fig:schedule}) is based on experience from \fermiLAT, the ComPair prototype, and a 2016 planning lab at GSFC. The five year mission is planned to launch in 2029.

%% file: cost.tex
\newpage
\section{Cost Estimates}\label{sec:costs}




An engineering design study for a similar mission concept was performed by a Goddard team in 2016. This mission had the same footprint, but fewer Silicon and CsI layers than AMEGO and did not have a CZT calorimeter. Cost estimates for this mission study were established with the CEMA/price-H and RAO teams at Goddard.

The cost category of AMEGO was established by scaling the price-H costs of the tracker and CsI calorimeter from the 2016 study and adding a bottoms up cost estimate for the CZT calorimeter based on vendor quotes and reasonable labor needs. Estimated spacecraft costs are also derived from the 2016 study using the upper end of the estimates from the price-H costing team at Goddard (note that this is a factor of two larger than the actual spacecraft cost for Fermi). The launch services assume a Falcon 9 launch to a low earth orbit. The science item covers development and operation of a science data center and a 5-year Guest Investigator program. The costs for the other mission elements are estimates derived by assuming average fractional mission costs for medium-sized missions. 

In this cost estimate, we assume that all support for AMEGO is from US Federal funds. However, we note that the expectation is that international contributions will provide a significant fraction of the payload.

The conclusion is that AMEGO comfortably fits in the Medium (Probe) cost category at $\sim$\$765M (\$995M including 30\% contingency). 

\begin{table}[htb!]
\small
\begin{tabular}{|p{5cm}|r|p{9.1cm}|}
\hline
\multirow{2}{*}{\bf WBS} & {\bf Cost} & \multirow{2}{*}{\bf Notes}\\ 
   & {\bf (\$M)} &  \\ 
\hline
Project Management & 61 & 8\% \\
Systems Engineering & 61  & 8\% \\
Safety and Mission Assurance & 38 & 5\% \\
Science & 86 & Includes science data center and 5-year GI program\\
Payload & 170 & Scaled instrument costs from 2016 study with additional bottoms-up estimate for CZT calorimeter\\
Spacecraft & 150 & Based on recent parametric estimates for similar spacecraft\\
Mission Operations & 122 & 16\% \\
Launch Services & 100 & Based on DSCOVR Falcon 9 launch costs \\
Ground Systems & 38 & 5\%\\
Systems Integration and Test & 38 & 5\% \\
\hline
Total & 765 & \\
Total with 30\% margin & 994 &\\ \hline
\end{tabular}
\caption{\small \it Estimated costs (FY\$17) per mission element.} \label{tab:wbs}
\end{table}

%% file: organization.tex
\section{Organization, Partnerships, and Current Status}\label{sec:org}


The AMEGO team is a large international group of $\sim$ 200 scientists in 15 countries, that brings together a large fraction of the space-based gamma-ray community with experience designing, building and operating gamma-ray telescopes (AGILE, ASCOT, BATSE, COMPTEL, COSI, EGRET, \fermiLAT, \textit{Fermi}-GBM, OSSE, etc.). The mission is currently led by Goddard Space Flight Center with significant contributions from other US and international partners.

The AMEGO team has significant overlap with the eAstroGAM consortium, who developed a similar mission concept for submission to the ESA medium class call.  European country agencies contributions to a future gamma-ray mission payload can apply to either AMEGO or eAstroGAM. In particular, we anticipate that these contributions will include Italians leveraging their expertise and experience to play a lead role in the Silicon tracker development, assembly, integration, test, and calibration; and French contributions to the tracker front-end electronics using the IDeF-X ASIC developed at CEA/Saclay. This ASIC has been selected for the STIX instrument of the Solar Orbiter mission. Letters of endorsement from international partner agencies can be provided upon request.

The AMEGO team across several institutions has an active program of hardware and software development to support the AMEGO mission. A NASA APRA award to GSFC and the Naval Research Laboratory (NRL) (PI: J. E. McEnery) supports the development of a prototype instrument with tests in beamline and a balloon flight. An additional APRA award to GSFC and Brookhaven National Laboratory (PI: D. J. Thompson) supports development of the CZT calorimeter technology. Development of the CsI calorimeter is supported by an APRA award to NRL (PI: J. E. Grove). Finally, software, simulations and analysis development support is provided via an APRA award to Berkeley (PI: A. Zoglauer). In addition to direct support from NASA for technology development, AMEGO also benefits from institutional support. GSFC and Argonne National Laboratory have provided internal resources to explore innovative Silicon detector technology. A proposal, with endorsement from the AMEGO team, has been submitted to the Japanese Society for the Promotion of Science to support an international collaboration in support of silicon detector and readout development and test for AMEGO.



\vspace{4mm}
\textbf{In Summary:} AMEGO will provide an exceptional view of the sky in the medium-energy gamma-ray domain. This is an under-explored energy range lies in a spectral gap of coverage between X-ray space telescopes and higher-energy gamma-ray instruments. 
AMEGO will bridge a key gap that to allow connections between the physical models of the sources with observations across the whole high-energy domain. 
Beyond that, this energy band offers unique breakthrough opportunities. 

The current state-of-the-art high-energy gamma-ray mission, NASA's Fermi Gamma-ray Space Telescope, turned 10 last year, reaching the mission lifetime goal without plans for a successor. AMEGO will address the anticipated absence of the energy and sky coverage provided by \fermiLAT by the end of the next decade. This absence will otherwise significantly diminish current capabilities in rapidly developing high-impact areas such as time-domain astrophysics, gravitational waves, and neutrino astronomy. 
We must start soon to minimize the impending loss of gamma-ray coverage and maximize the gain for the unique multi-wavelength and multimessenger instruments that are currently being upgraded or will become operational. The proposal team is strong, has first hand experience in building highly successful missions, and is full of enthusiasm to make AMEGO a timely reality.